# Measuring thermal contact resistances between metallic rods using laser spot heating and infrared thermography. Part 1.

Authors:
1.
Full name:
Thomas Lahens (corresponding author)
Affiliations:
1.Univ. Bordeaux, CNRS, Bordeaux INP, I2M, UMR 5295, F-33400, Talence, France
2.Arts et Metiers Institute of Technology, CNRS, Bordeaux INP, I2M, UMR 5295, F-33400 Talence, France
Institution address:
351 cours de la libération, 33400 Talence, France
Email address:
thomas.lahens@u-bordeaux.fr
2.
Full name:
Alain Sommier
Affiliations:
1.Univ. Bordeaux, CNRS, Bordeaux INP, I2M, UMR 5295, F-33400, Talence, France
2.Arts et Metiers Institute of Technology, CNRS, Bordeaux INP, I2M, UMR 5295, F-33400 Talence, France
Institution address:
351 cours de la libération, 33400 Talence, France
3.
Full name:
Marie-Marthe Groz
Affiliations:
1.Univ. Bordeaux, CNRS, Bordeaux INP, I2M, UMR 5295, F-33400, Talence, France
2.Arts et Metiers Institute of Technology, CNRS, Bordeaux INP, I2M, UMR 5295, F-33400 Talence, France
Institution address:
351 cours de la libération, 33400 Talence, France
4.
Full name:
Jean-Christophe Batsale
Affiliations:
1.Univ. Bordeaux, CNRS, Bordeaux INP, I2M, UMR 5295, F-33400, Talence, France
2.Arts et Metiers Institute of Technology, CNRS, Bordeaux INP, I2M, UMR 5295, F-33400 Talence, France
Institution address:
351 cours de la libération, 33400 Talence, France

**Abstract**
Estimation of thermal contact resistances between cylinders can be achieved using heating on the cross sections by a laser spot and measurement of the temperature response by IR thermography. This type of measurement makes it possible to characterize contacts in clusters of cylinders to simulate clusters of grains, or the properties of fibrous media in the transverse direction or, under certain conditions, to detect emerging thermal resistances (cracks perpendicular to the observation plane) by IR thermography. Here the thermal model is simplified because the temperature field in relaxation on either side of the resistance can be considered isothermal with a separation of transfers in the plane and along the thickness.
The mathematical model is then analytical (relaxation of a linear system) and the processing of the temperature field can be inversed thanks to the consideration of the logarithm of matrices, useful when studying the propagation of measurement noise.
The method proposed in the case of two or three cylinders is a first step and a validation for applications to the study of more complex media (large number of cylinders, granular media).

1. **Introduction**

Thermal characterization methods consisting of exciting the front face of a heterogeneous medium using a laser spot and analyzing the thermal response field on that same face, fall under a class of non-destructive thermal testing techniques.
The spot can be localized for measuring the diffusivity of locally homogeneous media in a pulsed regime [1-4] or periodic regime [5]. The models are analytical and allow for the determination of a thermal diffusivity tensor.
Studies of the response to a moving spot, commonly referred to as "flying-spot", also exist. The "flying-spot" has led to numerous applications, and even today, it continues to benefit from technological advances [6]. One of them is the detection of isolated through cracks in a conductive material [7-10]. In this case, the models and inversion methods are more complex. They often allow for localization and detection of the defect, but estimating thermal resistance is more challenging due to the complexity of the models and situations [11].
Few methods consider thermal effects in heterogeneous media with numerous through-contact resistances, such as granular media or cylinder bundles. In mechanics, the study of cylinder bundles often refers to the pioneering work of Schneebeli [12] to analyze stress distribution. Following these mechanical considerations, thermal studies using IR thermography often analyze the source terms generated at the interfaces between cylinders in cases where the stacking is subjected to periodic loading, for example in the elastic domain [13-14].
This paper proposes to complement these studies by measuring the thermal contact resistance between cylinders, without transient mechanical loading of the medium.
Two basic case studies of this type of system are presented, where modeling allows for a physical approach and generalization. First, the contact resistance between two cylinders is estimated, in which the model enables a nearly complete physical and analytical approach. Then the more general case of contact between three cylinders is presented, anticipating the study of granular media with a large number of grains or tubes.

Analytical models are well suited for industrial applications. For example, the method presented here can be used for in situ aerial inspection [15] of structures such as bridge suspension cables to evaluate parameters related to thermal contact resistance like strain distribution.

## 2. Experimental set-up

The "flying spot" set-up is represented on Figure 1. An $xyz$ stage allows measurements on samples with different shapes and dimensions. An infrared laser is directed to the sample thanks to a dichroic mirror. In this experiment, a 880 $nm$ direct diode laser from Innolas Laser Technology was used, along with an infrared camera from FLIR with a spectral range from 7.7 to 9.3 $\mu m$ (SC7000 series).

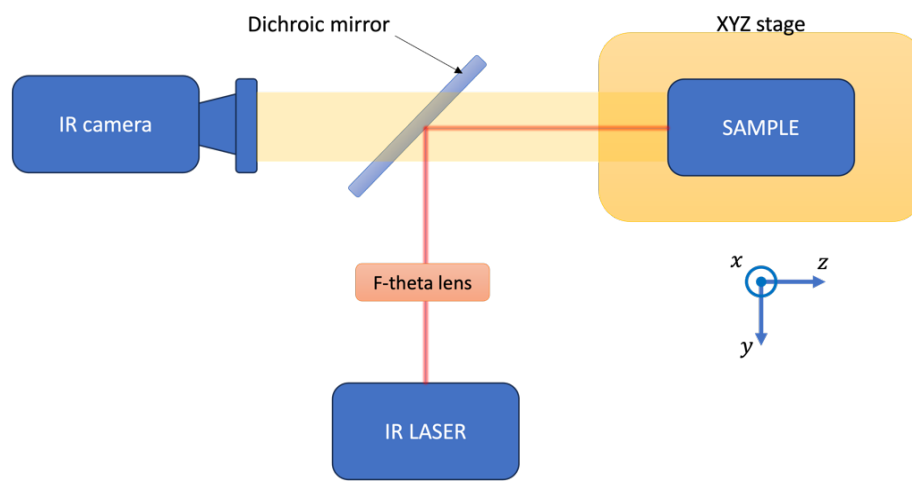

Figure 1: the "flying spot" experimental set-up.

## 3. Study of the contact between two cylinders

It is here considered that two metallic, solid cylinders of length $L$ and respective section $S_1$ and $S_2$ are in contact, as illustrated on Figure 2. They have the same thermal conductivity following the z-direction and the same heat capacity.

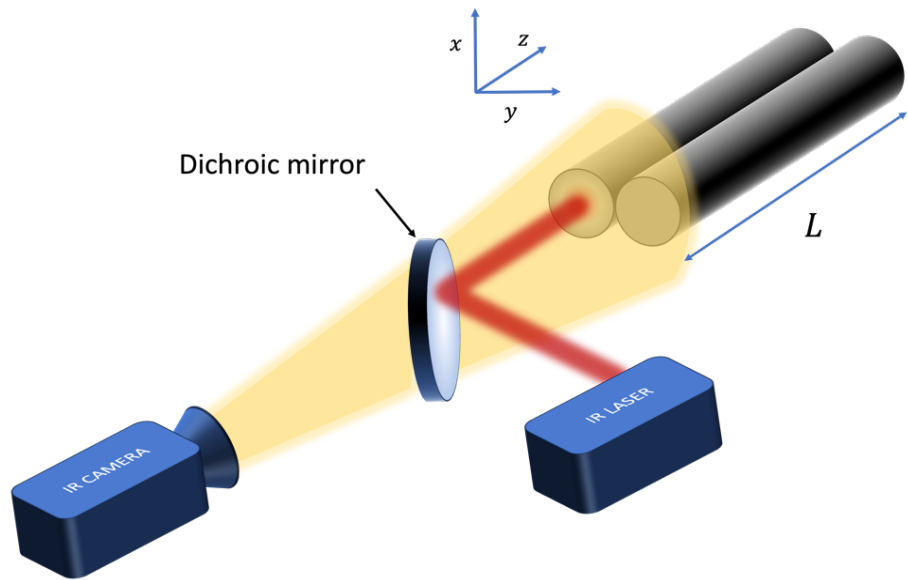

*Figure 2: scheme of the experiment with 2 cylinders.*

A heating pulse is applied on the front face ($z = 0$) of the cylinder 1 with a laser spot. The temperature responses of the front face $T_1(z = 0, t)$ and $T_2(z = 0, t)$ of cylinder 1 and cylinder 2 respectively, are measured with an infrared camera. For that purpose, they have been covered with a thin layer of black paint.

Although the diameter of the laser spot is of the order of $1 mm$ and the diameter of the cylinder is $D = 6 mm$, to evaluate the thermal contact resistance between the cylinders, the assumption is made that their perpendicular cross sections are isothermal. This hypothesis is valid if the in-plane ($xy$) diffusion time is negligible compared to the diffusion time along the z-axis. In a material with isotropic diffusivity, this is achieved by taking the length of the cylinders much larger than their diameter. The in-plane diffusion time can be estimated by $\sim D^2/a$, with $D$ the diameter of the cylinders and $a$ their thermal diffusivity. Because they are made of aluminum, this gives $\sim 300 \ ms$. Experimentally the characteristic time describing heat transfer between the cylinders is of the order of $10 \ s$, so the in-plane diffusion is much quicker compared to the time scale at which the cylinders exchange heat. The isothermal cross section hypothesis is then legitimate. To illustrate this result, Figure 3 shows cross-section thermograms at the end of the heat pulse, then $1,9 \ ms$ and $10 \ ms$ after. The temperature profile of the cross-section shows that after $10 ms$, they can already be considered isothermal.

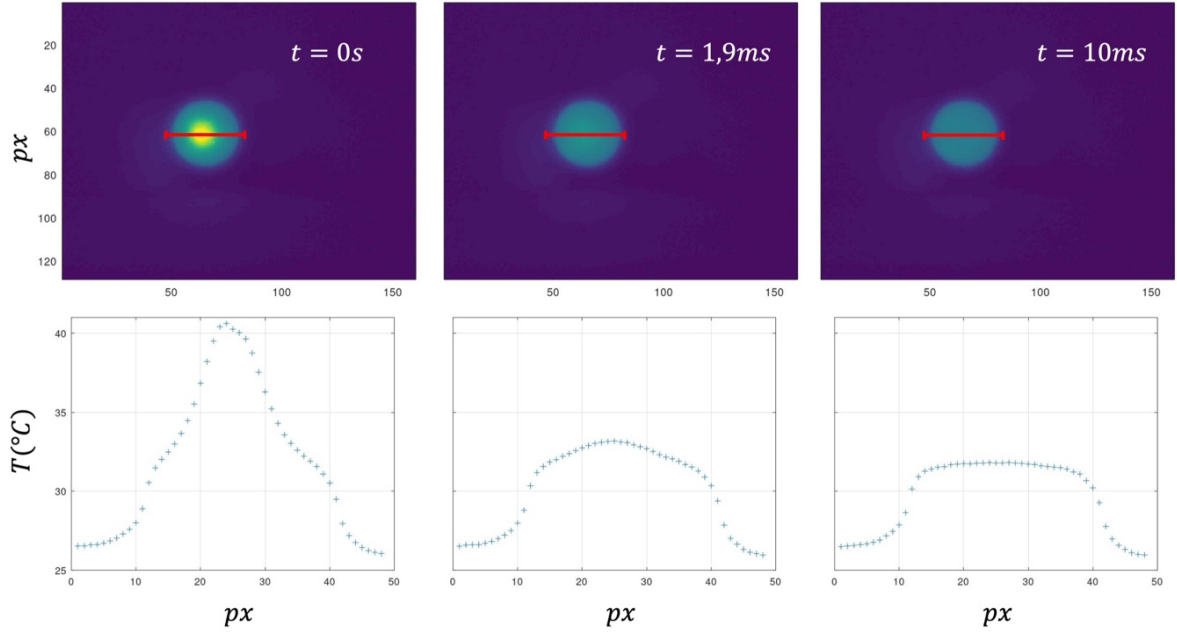

*Figure 3: thermograms of the front face of a cylinder (top) and the associated temperature profile (bottom) at three different times after the heat pulse.*

It is also considered that the lateral faces of the cylinders are adiabatic. This can be justified by calculating the characteristic time of the fin effect $\frac{\rho c D}{4h}$, $\rho c$ being the heat capacity and $h$ the convective heat transfer coefficient. In the case of $6mm$ diameter aluminum cylinders and considering typical values for $h$ ($5\ W.m^{-2}.K^{-1} < h < 25\ W.m^{-2}.K^{-1}$ for free convection), this gives a time of a few $100\ s$. This time being much larger than the transient behavior of heat transfer between cylinders (around $10\ s$), lateral convective losses can indeed be neglected.

Therefore, the heat equation is one-dimensional with a source term (heat pulse on the front face) and a leakage term (heat exchange between the cylinders). This gives the following system of equation [15]:

$$\begin{cases} \rho c \frac{\partial T_1}{\partial t} = \lambda \frac{\partial^2 T_1}{\partial z^2} + \frac{Q_1}{S_1}\delta(z)\delta(t) - \{leakage_{1\to 2}\} \\ \rho c \frac{\partial T_2}{\partial t} = \lambda \frac{\partial^2 T_2}{\partial z^2} + \frac{Q_2}{S_2}\delta(z)\delta(t) - \{leakage_{2\to 1}\} \end{cases} \quad (1)$$

$\rho c$ being the heat capacity in $J.K^{-1}.m^{-3}$, and $\lambda$ the thermal conductivity in $W.m^{-1}.K^{-1}$.

The term $\frac{Q_i}{S_i}\delta(z)\delta(t)$ account for the front face heat pulse, $Q_i$ being the energy carried by the heat pulse on the front face of cylinder $i$ in $J$, $S_i$ the cross-section area of cylinder $i$ in $m^2$, $\delta(z)$ and $\delta(t)$ representing the Dirac function in space and time, with respective units $m^{-1}$ and $s^{-1}$. For a more general consideration, it is here considered that the two cylinders can receive simultaneously two different levels of energy.

To express the leakage terms, both cylinders are treated as elastic bodies in non-adhesive normal contact, with smooth surfaces. Hence, the contact between the two parallel cylinders is not a line but a surface of area $l_c L$, with $l_c$ the width of the contact (Figure 4). In this paper, it will be assumed to be uniform along the length of the cylinders.

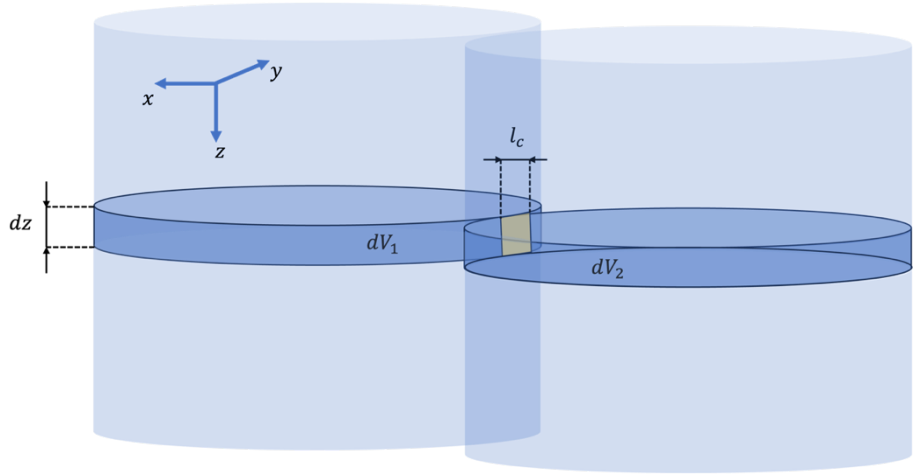

Figure 4: scheme of two elementary volumes, one for each cylinder, and the surface through which they exchange heat.

The one-dimensional heat equations of system (1), describe the evolution of the temperature $T_1(z,t)$ of an infinitely small volume $dV_1 = S_1 dz$ in contact with another infinitely small volume $dV_2 = S_2 dz$ at temperature $T_2(z,t)$. These two elementary volume exchange heat through the surface $S_c = l_c dz$, thus the heat flux $\Phi$ (in $W$, positive when going from 1 to 2) is expressed as:

$$\Phi = \frac{S_c}{R}(T_1 - T_2) \qquad (2)$$

With $R$ the thermal contact resistance in $m^2.K.W^{-1}$.

Each term of the heat equation, as it is written in system of equation (1), is in $W.m^{-3}$, as it describes the rate at which an elementary volume exchange heat relative to its volume. As a result, leakage term for cylinder 1 toward cylinder 2 can be expressed as:

$$\{leakage_{1 \to 2}\} = \frac{\Phi}{dV_1} = \frac{l_c}{RS_1}(T_1 - T_2) \qquad (3)$$

System of equation (1) then becomes:

$$\begin{cases} \rho c \dfrac{\partial T_1}{\partial t} = \lambda \dfrac{\partial^2 T_1}{\partial z^2} + \dfrac{Q_1}{S_1}\delta(z)\delta(t) - \dfrac{l_c}{RS_1}(T_1 - T_2) \\ \rho c \dfrac{\partial T_2}{\partial t} = \lambda \dfrac{\partial^2 T_2}{\partial z^2} + \dfrac{Q_2}{S_2}\delta(z)\delta(t) - \dfrac{l_c}{RS_2}(T_2 - T_1) \end{cases} \quad (4)$$

According to Hertz theory of contact [16], $l_c$ depends on the diameter of the cylinders, their respective Young's Modulus and Poisson's ratio and the force between the cylinders. However, within the framework of this paper, it is assumed that, either parts or all of these parameters are unknown. Therefore, the estimation method proposed in this work will be applied to the parameter:

$$R_l = \dfrac{R}{l_c} \quad (5)$$

This resistance, as it is defined, refers to a lineic conductance. Therefore, it is in units of $m.K.W^{-1}$. As stated above in the text, $l_c$ is assumed to be uniform along the axis of the cylinders so it will also be the case for $R_l$.

In the same way, the lineic heat capacities $C_1$ and $C_2$ of each cylinder are introduced:

$$C_i = \rho c S_i$$

It is in unit of $J.K^{-1}.m^{-1}$ and represents how much energy is needed to heat up by $1K$ the temperature of the corresponding cylinder per unit of length.

Finally, system of equation (1) simply becomes:

$$\begin{cases} \dfrac{\partial T_1}{\partial t} = a \dfrac{\partial^2 T_1}{\partial z^2} + \dfrac{Q_1}{C_1}\delta(z)\delta(t) - \dfrac{1}{R_l C_1}(T_1 - T_2) \\ \dfrac{\partial T_2}{\partial t} = a \dfrac{\partial^2 T_2}{\partial z^2} + \dfrac{Q_2}{C_2}\delta(z)\delta(t) - \dfrac{1}{R_l C_2}(T_2 - T_1) \end{cases} \quad (6)$$

With $a$ the thermal diffusivity of the material.

Heat fluxes are zero at the boundaries ($z = 0$ and $z = L$).

In practice, the cylinders were held together using a very thin layer of glue to make sure that a sufficient amount of energy was transferred from one to another. The glue is used to simulate a uniform, wider contact between the cylinders, thus eliminating the need for a complex mechanical system applying force. However, as shown on Figure 5, the thickness of this glue layer can be neglected given that the diameter of the cylinders is much larger.

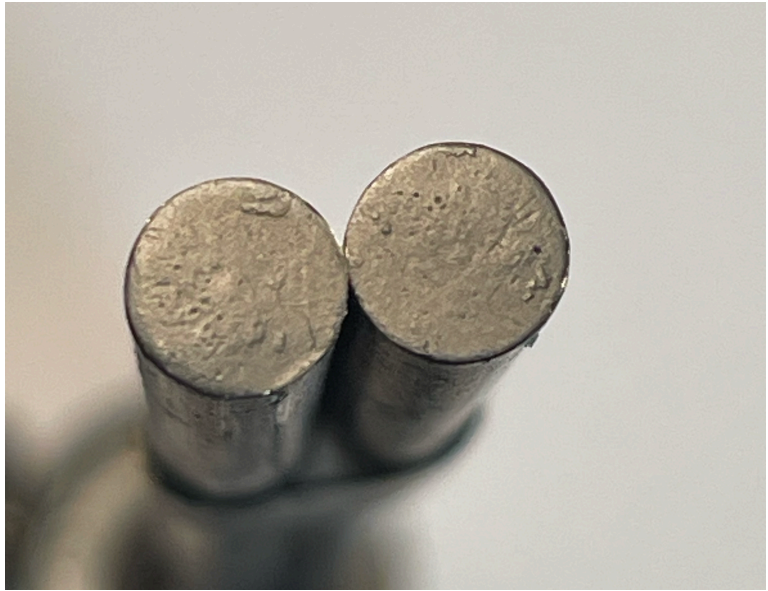

*Figure 5: picture of the front face of the cylinders pack showing that the glue thickness is negligeable compared to the dimensions of the system.*

Another way to consider the previous system is to use a double integral transform: a Laplace transform in time, and because of the adiabatic boundary conditions, a cosine Fourier transform on the z-variable, such that [17]:

$$\overline{\overline{T_i}}(\alpha_n, p) = \int_0^\infty \int_0^L \exp(-pt) \cos(\alpha_n z) T_i(z,t) dz dt \qquad (7)$$

With $\alpha_n = n\pi/L$,

Or the simple z-Cosine Fourier transform such that:

$$\overline{T_i}(\alpha_n, t) = \int_0^L \cos(\alpha_n z) T_i(z,t) dz \qquad (8)$$

System (6) becomes then under a more compact form:

$$\begin{cases} p\overline{\overline{T_1}} = -a\alpha_n^2 \overline{\overline{T_1}} - \dfrac{1}{R_l C_1}\left(\overline{\overline{T_1}} - \overline{\overline{T_2}}\right) + \dfrac{Q_1}{C_1} \\ p\overline{\overline{T_2}} = -a\alpha_n^2 \overline{\overline{T_2}} - \dfrac{1}{R_l C_2}\left(\overline{\overline{T_2}} - \overline{\overline{T_1}}\right) + \dfrac{Q_2}{C_2} \end{cases} \qquad (9)$$

The previous system can be presented in a matrix manner by:

$$\begin{bmatrix} \overline{\overline{T_1}} \\ \overline{\overline{T_2}} \end{bmatrix}_{p,\alpha_n} = ((p + a\alpha_n^2)[I] - [M])^{-1} \cdot \begin{bmatrix} Q_1/C_1 \\ Q_2/C_2 \end{bmatrix} \quad (10)$$

with $[I]$ the identity matrix and $[M]$ such that:

$$[M] = \begin{bmatrix} -1/(R_l C_1) & 1/(R_l C_1) \\ 1/(R_l C_2) & -1/(R_l C_2) \end{bmatrix}$$

The vector: $\left[\overline{\overline{T_1}}, \overline{\overline{T_2}}\right]^t$ is a function of $(p + a\alpha_n^2)$. The shift properties of the Laplace transform yields then, by considering the inverse Laplace transform in the cosine Fourier space:

$$\begin{bmatrix} \overline{T_1} \\ \overline{T_2} \end{bmatrix}_{\alpha_n,t} = \exp(-a\alpha_n^2 t) \; [expm([M]t)] \cdot \begin{bmatrix} Q_1/C_1 \\ Q_2/C_2 \end{bmatrix} \quad (11)$$

Where $[expm([M]t)]$ is the exponential of the matrix $[M]t$. This function is very efficiently implemented in most scientific programming languages, using a Padé approximant [18].

Another presentation in the $(z, t)$ space yields:

$$\begin{bmatrix} T_1 \\ T_2 \end{bmatrix}_{z,t} = \left( \frac{1}{L} + \frac{2}{L} \sum_{n=1}^{\infty} \exp(-a\alpha_n^2 t) \cos(\alpha_n z) \right) [expm([M]t)] \cdot \begin{bmatrix} Q_1/C_1 \\ Q_2/C_2 \end{bmatrix} \quad (12)$$

The thermal diffusion versus z direction can be here considered as separated from the diffusion between the cylinders.
For a 2X2 matrix, the study of the exponential of a matrix can be studied by considering the eigenvectors and eigenvalues of the $([M]t)$ matrix: the eigenvalues are 0, associated to the eigenvector $\begin{pmatrix} 1 \\ 1 \end{pmatrix}$, and $-\frac{(C_1+C_2)}{(R_l C_1 C_2)} t = -t/\tau$, associated to the eigenvector $\begin{pmatrix} C_2 \\ -C_1 \end{pmatrix}$.

Thanks to the properties of the exponential function of a matrix, we can write:

$$expm([M]t) = expm\left( \begin{bmatrix} 1 & C_2 \\ 1 & -C_1 \end{bmatrix} \cdot \begin{bmatrix} 0 & 0 \\ 0 & -t/\tau \end{bmatrix} \cdot \begin{bmatrix} 1 & C_2 \\ 1 & -C_1 \end{bmatrix}^{-1} \right)$$

$$= \begin{bmatrix} 1 & C_2 \\ 1 & -C_1 \end{bmatrix} \cdot \begin{bmatrix} 1 & 0 \\ 0 & \exp(-t/\tau) \end{bmatrix} \cdot \begin{bmatrix} 1 & C_2 \\ 1 & -C_1 \end{bmatrix}^{-1}$$

It yields a simple solution for the 2-cylinder problem separated from the z-diffusion:

$$\begin{bmatrix} T_1 \\ T_2 \end{bmatrix}_{z=0,t} = \begin{bmatrix} \dfrac{C_1}{C_1+C_2} + \dfrac{C_2 \exp\left(-\frac{t}{\tau}\right)}{C_1+C_2} & \dfrac{C_2}{C_1+C_2} - \dfrac{C_2 \exp\left(-\frac{t}{\tau}\right)}{C_1+C_2} \\ \dfrac{C_1}{C_1+C_2} - \dfrac{C_1 \exp\left(-\frac{t}{\tau}\right)}{C_1+C_2} & \dfrac{C_2}{C_1+C_2} + \dfrac{C_1 \exp\left(-\frac{t}{\tau}\right)}{C_1+C_2} \end{bmatrix} \cdot \begin{bmatrix} \dfrac{Q_1}{C_1} \\ \dfrac{Q_2}{C_2} \end{bmatrix} \quad (13)$$

Such analytical separated solutions are only possible if the cylinders are made of the same materials. More complete studies can be developed without separability.

If the length $L$ of the cylinder is considered infinite a simplified analytical expression of equation (12) when $z = 0$ (front face) is:

$$\begin{bmatrix} T_1 \\ T_2 \end{bmatrix}_{z=0,t} = (1/\sqrt{\pi a t})[expm([M]t)] \cdot \begin{bmatrix} Q_1/C_1 \\ Q_2/C_2 \end{bmatrix} \quad (14)$$

If the length $L$ is small, expression (14) becomes:

$$\begin{bmatrix} T_1 \\ T_2 \end{bmatrix}_{z=0,t} = \left(\frac{1}{L}\right)[expm([M]t)] \cdot \begin{bmatrix} Q_1/C_1 \\ Q_2/C_2 \end{bmatrix} \quad (15)$$

It is worth noting that, from system (9), $C_1\overline{\overline{T_1}} + C_2\overline{\overline{T_2}}$ is not depending on the heat diffusion between the cylinders, or not depending on $R$.

$$(p + a\alpha_n{}^2)\left(C_1\overline{\overline{T_1}} + C_2\overline{\overline{T_2}}\right) = Q_1 + Q_2 \quad (16)$$

Equation (16) can then be used to test the validity of this model. In the $(z, t)$ space, considering the cylinders are identical and only one of them is excited (as described on Figure 2), it yields:

$$T_1(z=0,t) + T_2(z=0,t) = \frac{1}{\sqrt{\pi a t}}\frac{Q}{C} \quad (17)$$

With $Q$ the energy of the laser pulse in Joule at the front face of the excited cylinder and $C = \rho c S$ the lineic capacity in $J.m^{-1}.K^{-1}$, $S$ being the cross-section of the cylinders.

Figure 6 shows the experimental sum of the front face temperatures in this case, on a logarithmic scale. It is fitted with a function $f(t) = K \times {1}/{\sqrt{t}}$. Front face temperatures were extracted by taking the mean value of the temperature on the cross-section of each cylinder.

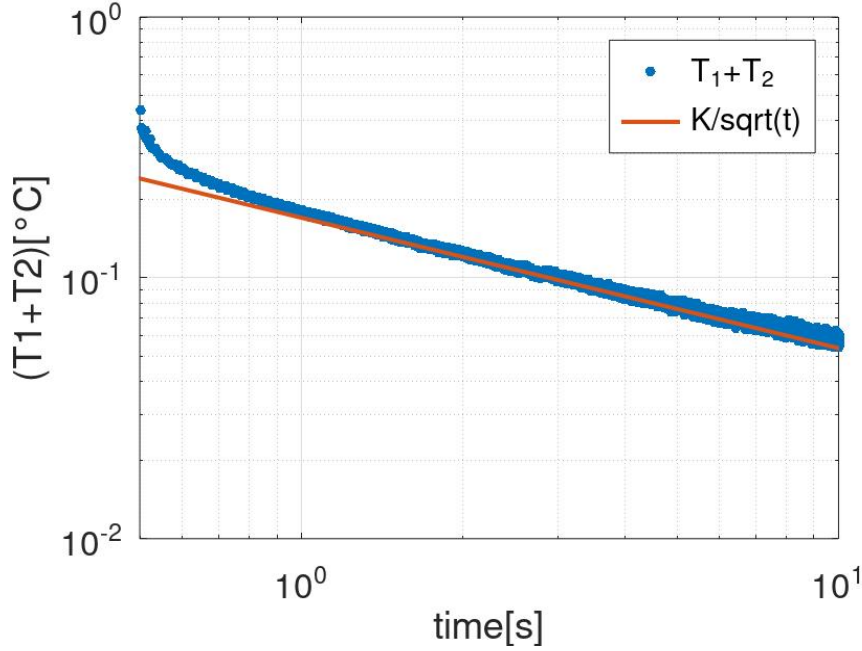

*Figure 6: experimental front face temperatures as a function of time on a logarithmic scale.*

At short timescale (before $1\ s$), the model does not quite fit the experiment because the heating pulse duration is $1\ s$ and, therefore, it cannot be described as a Dirac yet [18]. For times greater than $1\ s$, we can see that the blue curve is perfectly proportional to $1/\sqrt{t}$, thus confirming that the sum of the front face temperatures is indeed one-dimensional.

Given that the lineic capacities are known and equal ($C_1 = C_2$), the resistance between the cylinders can theoretically be calculated using the function $g(t)$ as:

$$g(t) = \ln\left(\frac{T_2 - T_1}{T_2 + T_1}\right) = -t/\tau + A \tag{18}$$

With $\tau = \frac{R_l C_1 C_2}{C_1 + C_2}$ and $A$ being constant.

The expression of function $g(t)$ can be calculated using equation (13) and (17). It is particularly convenient because, as stated earlier in the text, in the expression of the front face temperatures, the in-plane and in-depth diffusion are separated. Consequently, the ratio of the difference over the sum of the front face temperatures is only dependent on the in-plane transfer, which is characterized by the time constant $\tau$, proportional to $R_l$.

$g(t)$ is plotted on Figure 7 using the experimental temperatures and is fitted with a linear curve.

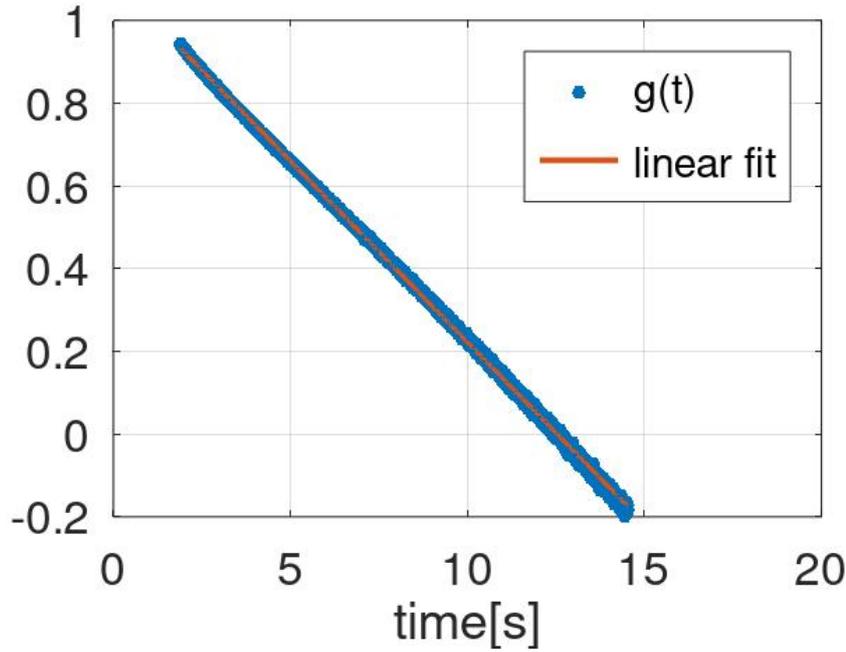

*Figure 7: function g(t) used for the estimation of the characteristic time describing heat exchange between the cylinders.*

Estimating $\tau$ with the least squares criterion and using typical values for the properties of aluminum ($\rho = 2700\ kg.m^{-3}$ and $c = 897\ J.K^{-1}.kg^{-1}$) yields:

$$R_l = (3333 \pm 17).10^{-4}\ m.K.W^{-1}$$

Because this is in units of $m.K.W^{-1}$, it is complicated to compare with typical values. However, because the cylinders are held together with a very thin layer of glue, the width of the contact $l_c$ can be assumed to be of the order of $1mm$. This gives a contact resistance:

$$R_{2rod} \sim (3{,}33 \pm 0{,}017).10^{-4}\ m^2.K.W^{-1}$$

This value falls in the same range as typical contact resistances between metals under low load. Besides, this estimation is highly robust to measurement noise, as showed by the confidence interval on $R_{2rod}$.

The model presented here gives promising results in the 2-cylinder case. In the next section, the 3-cylinder case is explored using a matrix formalism to evaluate the thermal contact resistances.

## 4. Study of the contact between three cylinders

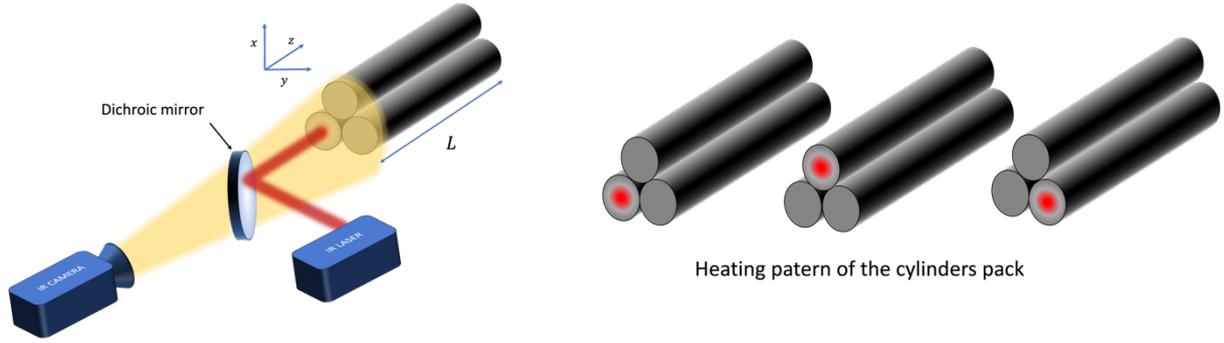

*Figure 8: experimental set-up for the 3-cylinder case.*

The experimental setup remains unchanged, except for the number of cylinders, as described on Figure 8. However, the experimental method is different: a heat pulse is applied on the front face ($z = 0$) of each of them successively and the front face temperatures are recorded with an infrared camera. The delay between the heat pulses is adjusted so that the system has totally relaxed.

The temperature of cylinder $i$ when cylinder $j$ has been heated will be written $T_{ij}(z,t)$. Because the physical hypotheses are identical to the 2-cylinder case, it is possible to write, similarly to equation (10):

$$\begin{bmatrix} \overline{\overline{T_{11}}} & \overline{\overline{T_{12}}} & \overline{\overline{T_{13}}} \\ \overline{\overline{T_{21}}} & \overline{\overline{T_{22}}} & \overline{\overline{T_{23}}} \\ \overline{\overline{T_{31}}} & \overline{\overline{T_{32}}} & \overline{\overline{T_{33}}} \end{bmatrix} = \left((p + a\alpha_n^2)[I] - [M]\right)^{-1} \cdot \begin{bmatrix} Q_1/C_1 & 0 & 0 \\ 0 & Q_2/C_2 & 0 \\ 0 & 0 & Q_3/C_3 \end{bmatrix} \quad (19)$$

However, matrix $[M]$ is expressed as:

$$[M] = \begin{bmatrix} -\dfrac{1}{R_{12}C_1} - \dfrac{1}{R_{13}C_1} & \dfrac{1}{R_{12}C_2} & \dfrac{1}{R_{13}C_3} \\ \dfrac{1}{R_{12}C_1} & -\dfrac{1}{R_{12}C_2} - \dfrac{1}{R_{23}C_2} & \dfrac{1}{R_{23}C_3} \\ \dfrac{1}{R_{13}C_1} & \dfrac{1}{R_{23}C_2} & -\dfrac{1}{R_{13}C_3} - \dfrac{1}{R_{23}C_3} \end{bmatrix}$$

$R_{12}$, $R_{23}$ and $R_{13}$ being the thermal resistances (in the way they are defined in equation (5)) respectively between cylinders: 1 and 2, 2 and 3, 1 and 3.

$Q_i$ and $C_i$ represent respectively the energy of the laser pulse when cylinder $i$ is being heated and its lineic capacity.

Once again, thanks to the shift properties of the Laplace transform and considering the inverse Laplace transform in the cosine Fourier space, equation (19) can be written in the $(z, t)$ space:

$$\begin{bmatrix} T_{11} & T_{12} & T_{13} \\ T_{21} & T_{22} & T_{23} \\ T_{31} & T_{32} & T_{33} \end{bmatrix}_{z,t} = \left( \frac{1}{L} + \frac{2}{L} \sum_{n=1}^{\infty} \exp(-a\alpha_n^2 t) \cos(\alpha_n z) \right) [expm([M]t)] . \begin{bmatrix} Q_1/C_1 & 0 & 0 \\ 0 & Q_2/C_2 & 0 \\ 0 & 0 & Q_3/C_3 \end{bmatrix} \quad (20)$$

As described in Figure 8, the experimental setup uses an infrared camera looking at the cross-section of the cylinders at $z = 0$. This means that only the front face temperatures are measured. However, initial and boundary conditions give:

$$\begin{bmatrix} Q_1/C_1 & 0 & 0 \\ 0 & Q_2/C_2 & 0 \\ 0 & 0 & Q_3/C_3 \end{bmatrix} = \begin{bmatrix} T_{11} & 0 & 0 \\ 0 & T_{22} & 0 \\ 0 & 0 & T_{33} \end{bmatrix}_{z=0, t=0}$$

Besides, because the length of the cylinders is much larger than their diameter, equation (20) can be approximated at $z = 0$ as follows:

$$\begin{bmatrix} T_{11} & T_{12} & T_{13} \\ T_{21} & T_{22} & T_{23} \\ T_{31} & T_{32} & T_{33} \end{bmatrix}_{z=0,t} = \frac{1}{\sqrt{\pi a t}} expm([M]t) . \begin{bmatrix} T_{11} & 0 & 0 \\ 0 & T_{22} & 0 \\ 0 & 0 & T_{33} \end{bmatrix}_{z=0, t=0} \quad (21)$$

Therefore, all thermal contact resistances can be evaluated through the identification of the elements in matrix $[M]t$ using the equation:

$$[M]t = logm \left( \sqrt{\pi a t} \begin{bmatrix} T_{11} & T_{12} & T_{13} \\ T_{21} & T_{22} & T_{23} \\ T_{31} & T_{32} & T_{33} \end{bmatrix}_{z=0,t} . \begin{bmatrix} T_{11} & 0 & 0 \\ 0 & T_{22} & 0 \\ 0 & 0 & T_{33} \end{bmatrix}_{z=0, t=0}^{-1} \right) \quad (22)$$

With *logm* being the logarithm of a matrix, or the inverse function of the exponential of a matrix. Similarly to the exponential of a matrix, the logarithm of a matrix is implemented in most scientific programming languages.

It is worth noting that, for this method to work, all cylinders must be excited individually.

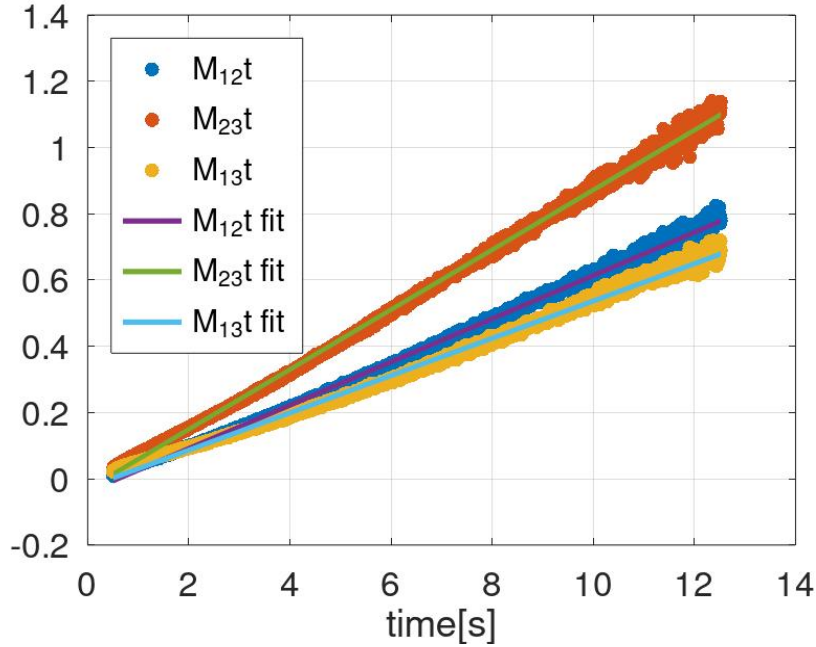

*Figure 9: plots of the elements of matrix [M]t as a function of time used for the estimation of all thermal contact resistances.*

On Figure 9, the elements $M_{12}t, M_{13}t$ and $M_{23}t$ of matrix $[M]t$ are plotted as a function of time, using experimental data in equation (22), as well as a linear regression curve for each of them. According to the definition of $[M]$, the slope of each curve gives $1/RC$.

Because in this case lineic capacities are known, experimental values for the thermal contact resistances, as defined in equation (5), are estimated using the least squares method:

$$R_{12} = (2238 \pm 75)10^{-4} \; m.K.W^{-1}$$
$$R_{13} = (2594 \pm 141)10^{-4} \; m.K.W^{-1}$$
$$R_{23} = (1617 \pm 50)10^{-4} \; m.K.W^{-1}$$

Again, assuming the width of the contact is around $1mm$, in units of $m^2.K.W^{-1}$, this gives:

$$R'_{12} \sim (2{,}238 \pm 0{,}075).10^{-4} m^2.K.W^{-1}$$
$$R'_{13} \sim (2{,}594 \pm 0{,}141).10^{-4} m^2.K.W^{-1}$$
$$R'_{23} \sim (1{,}617 \pm 0{,}05).10^{-4} m^2.K.W^{-1}$$

Which, again, falls in the range of classical values for metallic surfaces in contact under low load. Besides these results are consistent with those reported in the literature [11]. Discrepancies from the model account for slightly non-uniform contact along the axis of the cylinders.

## 5. Conclusion and perspectives

This paper examines heat diffusion in a pack of cylinders. While the problem is initially complex, it is shown that assuming the cross-section of the cylinders are isothermal and neglecting lateral convective losses, thermal contact resistances can be identified in both the 2 and 3-cylinder case. For the cross-section of the cylinders to be considered isothermal, the characteristic time for heat transfer between the tubes must be much larger than the in-plane diffusion time. This is true for highly diffusive materials. If the cylinders exchange heat much slower than what is presented here, convective losses can't be neglected. However, they can be calculated by taking the sum of all the front face temperatures and taken into account in the estimation of the thermal contact resistances.

The results shown for the 2-cylinder case demonstrate the physical approach is coherent, while the results for the 3-cylinder case show that, thanks to the already existing algorithms calculating matrix exponentials and logarithms, calculation of the matrix of thermal contact resistances are efficient and robust to measurement noise. Only noise measurement on the experimental temperatures has been taken into account to evaluate confidence intervals. Uncertainties on the camera calibration or the physical parameters such as aluminum heat capacity have been neglected. The experimental estimations are in agreement with those found in the literature. However, it is worth noting that they are only made possible because in the expression of the temperature field, the in-plane diffusion was separable from the in-depth diffusion.

Upcoming work includes investigation on the detection of non-uniform contact between cylinders. More precisely, part 2 of this work will focus on the special case of a rupture in the contact between the cylinders, in the 2-cylinder case. The method presented in this paper can also be extended to packs of $N$ cylinders. This last point opens the way to the study of thermal properties of granular media through IR thermography and the estimation of more general heterogeneous material properties.

ACKNOWLEDGMENTS:

The authors gratefully acknowledge the financial support of FranceRelance and France2030 through the SMART CABLE project, and the European Union through the NextGenEU project.